\documentclass{elsart}

\usepackage{graphics}
\usepackage{graphicx}
\usepackage{dcolumn}
\usepackage{bm}

\usepackage{latexsym}
\usepackage{bbm}                      
\usepackage{ifthen}                   
\usepackage{exscale}                  
\usepackage[intlimits]{amsmath}       
\usepackage{amsfonts}
\usepackage{amssymb,amscd}
\usepackage[dvips]{epsfig}                   
\usepackage{array}
\usepackage{wrapfig}

\newcommand*{\bea}{\begin{eqnarray}}
\newcommand*{\eea}{\end{eqnarray}}
\newcommand*{\be}{\begin{equation}}
\newcommand*{\ee}{\end{equation}}
\newcommand*{\pd}{\partial}

\newcommand*{\pref}[1]{(\ref{#1})}

\newcommand*{\tr}{\mathrm{tr}}

\newcommand*{\Tr}{\mathrm{Tr}}

\begin{document}

\begin{frontmatter}

\title{Infrared-suppressed gluon propagator \\
        in 4d Yang-Mills theory \\ in a Landau-like gauge}

\author[a]{Attilio Cucchieri}\ead{attilio@ifsc.usp.br}
\author[a,b]{Axel Maas}\ead{axelmaas@web.de}
\author[a]{Tereza Mendes}\ead{mendes@ifsc.usp.br}

\address[a]{Instituto de F\'\i sica de S\~ao Carlos, Universidade de S\~ao Paulo,
               Caixa Postal 369, 13560-970 S\~ao Carlos, SP, Brazil}
\address[b]{Gesellschaft f\"ur Schwerionenforschung mbH, Planckstra{\ss}e 1,
                    D-64291 Darmstadt, Germany}

\date{\today}


\begin{abstract}
The infrared behavior of the gluon propagator is directly related to confinement in QCD.
Indeed, the Gribov-Zwanziger scenario of confinement predicts an infrared vanishing
(transverse) gluon propagator in Landau-like gauges, implying violation of
reflection positivity and gluon confinement. Finite-volume effects make
it very difficult to observe (in the minimal Landau gauge) an infrared suppressed
gluon propagator in lattice simulations of the four-dimensional case.
Here we report results for the $SU(2)$ gluon propagator in a gauge that interpolates
between the minimal Landau gauge (for gauge parameter $\lambda$ equal to 1) and 
the minimal Coulomb gauge (corresponding to $\lambda = 0$).
For small values of $\lambda$ we find that the spatially-transverse
gluon propagator $D^\tr(0,|\vec p|)$, considered as a function of the spatial momenta
$|\vec p|$, is clearly infrared suppressed. This result is in agreement with the
Gribov-Zwanziger scenario and with previous numerical results in the minimal
Coulomb gauge. We also discuss the nature of the limit $\lambda \to 0$ (complete
Coulomb gauge) and its relation to the standard Coulomb gauge ($\lambda = 0$).
Our findings are corroborated by
similar results in the three-dimensional case, where the infrared suppression is observed
for all considered values of $\lambda$.
\end{abstract}

\begin{keyword}
Yang-Mills theory \sep
Gluon propagator \sep
Confinement \sep
Interpolating gauge
\PACS 11.15.Ha
\sep  12.38.Aw
\sep  14.70.Dj
\end{keyword}
\end{frontmatter}


\section{Introduction}

The infrared behavior of the gluon propagator is linked to the confinement of
gluons.\cite{Alkofer:2000wg} In particular, the confinement scenario of Gribov and
Zwanziger\cite{Gribov,gribov1,gribov2,Zwanziger:1993dh,Zwanziger,zwanziger1,Zwanziger:2003cf}
predicts a (transverse) gluon propagator vanishing at zero (Euclidean) momentum in Landau
gauge and in Landau-like gauges (or $\lambda$-gauges). 
The latter class refers to gauges interpolating between the
Landau and the Coulomb gauge\cite{Baulieu:1998kx}, with a gauge condition (in
$d$ dimensions) given by
\be
\lambda \, \pd_0 A_0^a \, + \,\pd_1 A_1^a \, + \, \ldots \, + \,\pd_{d-1} A_{d-1}^a
    \,= \, 0 \; ,
\label{gcinterpol}
\ee
where the gauge parameter $\lambda$ is between 1 and 0.
Let us recall that an infrared (IR) null gluon propagator has
far-reaching consequences. Indeed, such a particle cannot have a positive
semi-definite spectral function\cite{Oehme,oehme1,oehme2,oehme3} or, as a consequence, a K\"allen-Lehmann
representation. This is regarded as one possible manifestation of confinement,\cite{Alkofer:2000wg,Mandula:1999nj,Alkofer:2003jj} when considering
Euclidean correlation functions.\footnote{Thus,
it is a sufficient condition for gluon confinement.\cite{Alkofer:2000wg,Alkofer:2003jj}}

The question of whether the Landau-gauge gluon propagator is indeed null at zero momentum
is a long-standing one. Various continuum methods, based on functional approaches, yield
a vanishing gluon propagator.\cite{Alkofer:2000wg,Zwanziger,zwanziger1,Zwanziger:2003cf,vonSmekal,Lerche:2002ep,vs1,vs2,vs3,vs4,vs5,vs6,vs7,Fischer:2006vf}
This result is rather tightly constrained,\cite{Fischer:2006vf} i.e.\
it seems to be the only possible solution satisfying both Dyson-Schwinger equations
and functional renormalization-group equations. At the same time, lattice calculations
in four dimensions have obtained an IR-suppressed Landau-gauge gluon propagator $D(p)$ only
when using strongly-asymmetric lattices\cite{Silva:2005hb} or a coupling constant in the
strong-coupling regime.\cite{Cucchieri:1997fy} For lattice couplings in the scaling region
and using symmetric lattices, one finds for the Landau-gauge gluon propagator an
increase slower than $1/p^2$ as one approaches the IR region\cite{irgluon,ir1,ir2,ir3}, 
with a finite value for
$p = 0$.\cite{large,large1,large2,Cucchieri:2006xi} The fact that a propagator decreasing at small momenta
is not observed in the 4d case, even for volumes of almost (10 fm)$^4$,\cite{Cucchieri:2006xi}
is probably due to very strong finite-size effects.\cite{large,large1,large2,Cucchieri:2006xi,sternbeck,sternbeck1,fv,fv1}
This assumption is supported by numerical results in the 3d case, where much larger lattice
sides are accessible. In this case, there is substantial evidence for an IR-suppressed
gluon propagator in Landau gauge,\cite{Cucchieri:1999sz,Cucchieri:2001tw,Cucchieri:2003di}
in agreement with continuum calculations.\cite{Zwanziger,Maas:2004se} However, also in
this case, a reliable extrapolation of $D(0)$ to the infinite-volume limit is still
lacking.\cite{Cucchieri:2003di} Finally, let us recall that lattice Landau calculations\cite{Langfeld,langfeld1,sternbeck,sternbeck1} have also obtained
direct evidence for the non-positivity of the gluon spectral function, both in the three- and
in the four-dimensional cases.

The Gribov-Zwanziger confinement scenario applies also to Coulomb gauge.\cite{Gribov,gribov1,gribov2,DanCoulomb,dc1,dc2,reviews,reviews1} 
In this case it is important to observe that the standard
Coulomb-gauge-fixing condition $\, \pd_1 A_1^a \, + \, \ldots \, + \,\pd_{d-1} A_{d-1}^a
\,= \, 0 \, $ 
is not a complete one, due to the residual gauge degrees of freedom $g(t)$. On the
other hand, a possible complete Coulomb gauge condition\cite{Baulieu:1998kx} can be obtained
using the class of gauges defined in \pref{gcinterpol}. Indeed, the parameter $\lambda$
interpolates between the Landau ($\lambda = 1$) and a complete Coulomb gauge, corresponding
to the limit\footnote{Clearly, since the gauge fixing \pref{gcinterpol} is complete for any
$\lambda \neq 0$, it is also a complete one when considering the limit $\lambda \to 0$.\cite{Baulieu:1998kx}} 
$\lambda \to 0$. Therefore, the complete
Coulomb gauge condition is, by definition, a smooth limiting case of the interpolating gauge
\pref{gcinterpol} while, of course, this is not the case for the standard Coulomb condition
($\lambda = 0$). Let us recall that the gauge condition \pref{gcinterpol} above can be
obtained by minimizing the (lattice) functional\footnote{A similar functional can be defined
in the continuum.}
\be
{\mathcal E}[g] \,=\, - \,\Tr \sum_{x} \; \left[  \, \lambda \, U_0(x) \,+ \,
      \sum_{i=1}^{d-1} \, U_i(x) \, \right] \; ,
\label{eq:defElambda}
\ee
where $U_\mu(x)$ indicates a lattice link variable in the $\mu$ direction. Then, the
limiting case $\lambda \to 0$ corresponds to minimizing the following two
functionals\cite{Cucchieri:2000gu}
\bea
{\mathcal E}_{\mbox{hor}}[g(\vec x)] &=& - \,\Tr \sum_{x} \; \sum_{i=1}^{d-1} \, U_i(x)
                                                                 \label{ECoulomb} \\
{\mathcal E}_{\mbox{ver}}[g(t)] &=& - \,\Tr \sum_{x} \; U_0(x) \; .  \label{Eresidual}
\eea
The minimization of the first functional is equivalent to a Landau gauge condition fixed on
each time slice, using $g(\vec x)$ gauge transformations, i.e.\ it corresponds to the
standard (incomplete) Coulomb gauge. The minimization of the second functional,
considering only $g(t)$ gauge transformations, provides additional constraints, necessary
to eliminate the residual gauge degrees of freedom. Note that we can also write 
$ \,{\mathcal E}_{\mbox{ver}}[g(t)] = - \,\Tr \sum_{t} \; Q_0(t) \,$,
with $\, Q_0(t) = \sum_{\vec x} \; U_0(t,\vec x) \,$.
Then, the minimization of ${\mathcal E}_{\mbox{ver}}[g(t)]$ is like a one-dimensional
Landau gauge fixing.
Of course, quantities defined in terms of spatial link variables $U_i(x)$
are not affected by the residual gauge condition obtained by minimizing the functional
${\mathcal E}_{\mbox{ver}}[g(t)]$.

Numerical studies in minimal Coulomb gauge have shown [for the $SU(2)$ case in 4d] that the
instantaneous transverse gluon propagator $D^{tr}(\vec p)$ is indeed suppressed in the IR 
limit.\cite{Cucchieri:2000gu,Cucchieri,cucchieri1,otherCoulomb,oc1} Also, in the infinite-volume
limit, it has been found\cite{Cucchieri:2000gu,Cucchieri,cucchieri1} that $D^{tr}(\vec p)$ is well
described by a Gribov-like propagator with a pair of purely imaginary poles
$m^2 = \pm i y$. These results are in agreement with the Gribov-Zwanziger confinement 
scenario.\cite{Gribov,gribov1,gribov2,DanCoulomb,dc1,dc2} The fact that, for a given lattice size $L$
(in fm), one sees an IR-suppressed transverse gluon propagator
in 4d-Coulomb gauge and in 3d-Landau gauge
but not in 4d-Landau gauge may be related to a quantitatively different IR suppression
in the two cases.
Indeed, functional methods\cite{Zwanziger,Maas:2004se,Coulomb,coulomb1,coulomb2,coulomb3} predict a stronger suppression
in 4d-Coulomb gauge and in 3d-Landau gauge than in 4d-Landau gauge, i.e.\ the so-called IR
gluon exponent $\alpha_D$ should be larger for the 4d-Coulomb and the 3d-Landau cases.

One should note that, for any non-zero value of $\lambda$, the gauge condition
\pref{gcinterpol} is essentially a {\em deformed} Landau gauge, i.e.\ the IR
exponents of the propagators do not depend on $\lambda$\cite{Fischer:2005qe}. 
In particular, calculations using Dyson-Schwinger equations suggest\cite{Fischer:2005qe} 
that, at momenta sufficiently small compared to a separation momenta $p_s$,
the transverse gluon propagator behaves as in Landau gauge, i.e.\ all
Lorentz and color components of the gluon propagator vanish at zero four-momentum.
However, the limit $\lambda \to 0$ can also be thought of as sending to zero the momentum
$p_s$, which separates Coulomb-gauge-like from Landau-gauge-like behavior.
Indeed, for any finite value $\lambda \neq 0$ and given the Landau gauge condition
$ \pd_{\mu} A_{\mu}^a = 0 $, we can obtain the gauge condition
\be
\pd_0 A_0^a \, + \lambda^{-1} \left[ \,\pd_1 A_1^a \, + \, 
                  \ldots \, + \,\pd_{d-1} A_{d-1}^a \, \right] = \, 0 \;
\ee
by using the rescaling\footnote{In Ref.~\cite{Fischer:2005qe}, a similar rescaling 
was used to show that Dyson-Schwinger equations for $\lambda$-gauges are equivalent 
to the Landau case for all $\lambda\neq 0$ .} $x_i \to x_i / \lambda$ for $i \neq 0$. 
This implies, in momentum space, the rescaling $p_i \to \lambda \, p_i$. Thus, if we 
consider only spatial momenta, we find\footnote{Of course, this simple explanation 
is correct only at tree-level and can be modified by the renormalization of
$\lambda$.} that $p_s$ is rescaled to $\lambda \, p_s$ and goes to zero when
$\lambda \to 0$. As a consequence, one should expect that, for very small values
of $\lambda$, all correlation functions would show a Coulomb-like behavior
for momenta $p > p_s$, with $p_s$ very small. In particular, the correlation
function that corresponds to the transverse (instantaneous) gluon propagator in
Coulomb gauge should become more and more IR suppressed as the parameter
$\lambda$ approaches zero. Investigating whether this is the case is the aim of this work.


\section{Numerical results}

Following Eqs.\ (8) and (9) of Ref.~\cite{Fischer:2005qe}, we can consider on
the lattice the three-dimensionally transverse gluonic correlation function
\be
D^\tr(p_0,|\vec p|) \; = \; \left(\delta_{ij} - \frac{p_i p_j}{{\vec p\,}^2}\right)
     \frac{ < A_i^a(p)  A_j^a(-p) > }{(N_c^2-1) (d-2) V} \label{dtr} \; .
\ee
At zero four-momentum one has 
\be
D^\tr(0) \; = \; \delta_{ij} \frac{ < A_i^a(0)  A_j^a(0) > }{(N_c^2-1) (d-1) V}
       \label{dtr0} \; .
\ee
Here, $N_c$ is the number of colors, $V$ is the $d$-dimensional lattice volume,
Lorentz indices $i, j$ are summed only over the $d-1$ spatial directions and
$A_i^a(p)$ is the $d$-dimensional Fourier transform of the gluon field.
Let us recall that, by considering only spatial momenta (i.e.\ $p_0 = 0$), the function
$\,D^\tr(0,|\vec p|)\,$ is predicted to be IR suppressed --- and vanishing at zero
momentum --- for all non-zero values of $\lambda$ in three and in four dimensions.\cite{Fischer:2005qe} 
Also, when $\lambda$ is null,
the above definition yields the instantaneous part of the three-dimensional transverse
gluon propagator\cite{Cucchieri:2000gu}
\be
D^{\mathrm{inst}}(|\vec p|) \; = \; \sum_t 
     \left(\delta_{ij} - \frac{p_i p_j}{{\vec p\,}^2}\right)
     \frac{ < A_i^a(t,\vec p)  A_j^a(t,-\vec p) > }{(N_c^2-1) (d-2) V} \label{dinst} \; ,
\ee
where now the Fourier transform of the gluon field is evaluated for each time slice.
Indeed, when $\lambda$ is null, the gauge transformations $g(t)$ are independent of
the gauge transformation $g(\vec x)$ and the
two sets of transformations commute. Then, the terms in Eqs.\ \pref{dtr} that depend
explicitly on $g(t)$ are averaged to zero and one is left with the expression above.
Note that this explanation is valid whether the residual gauge freedom $g(t)$ is 
fixed or not.

Here we evaluate numerically $D^\tr(0,|\vec p|)$ as a function of $|\vec p|$ for $SU(2)$
Yang-Mills theory in three and in four dimensions for several values of the parameter $\lambda$. 
Details of the simulations can be found in.\cite{Cucchieri:2003di,Cucchieri:2006tf}
Let us note that for $\lambda \neq 1$ the numerical gauge fixing is
very similar to the usual Landau gauge fixing.\cite{Cucchieri:2001tw} On the
other hand, when $\lambda$ goes to zero one sees\cite{Cucchieri:1998ta} that more
iterations are needed in order to satisfy a given numerical accuracy for the gauge fixing,
in agreement with a recent analytic study.\cite{Cucchieri:2003fb}
This problem can be partially reduced by adding some extra gauge fixing
sweeps in the $\mu=0$ direction for each iteration of the gauge-fixing algorithm, i.e.\
by considering gauge transformations $g(t)$ that depend only on the $\mu=0$ component of $x$.
Finally, we did not consider here possible systematic effects related to the breaking
of rotational symmetry or to the existence of Gribov copies. The former type of effects
can be parameterized by\cite{Ma:1999kn} $a^2 p^{[4]}$, where $a$ is the lattice
spacing and $p^{[4]} = \sum_{\mu} p_{\mu}^4$. Therefore, this type of effects are
not expected to play a significant role in the IR region, considered here.
As for the latter type of effects, in the infinite-volume limit, averages taken over 
configurations belonging to the so-called Gribov region $\Omega$ should coincide\cite{Zwanziger:2003cf}
with averages obtained by restricting the functional integral to the so-called fundamental
modular region $\Gamma$, whose interior is free of Gribov copies.

\begin{figure}
\begin{center}
\includegraphics[width=0.7\linewidth]{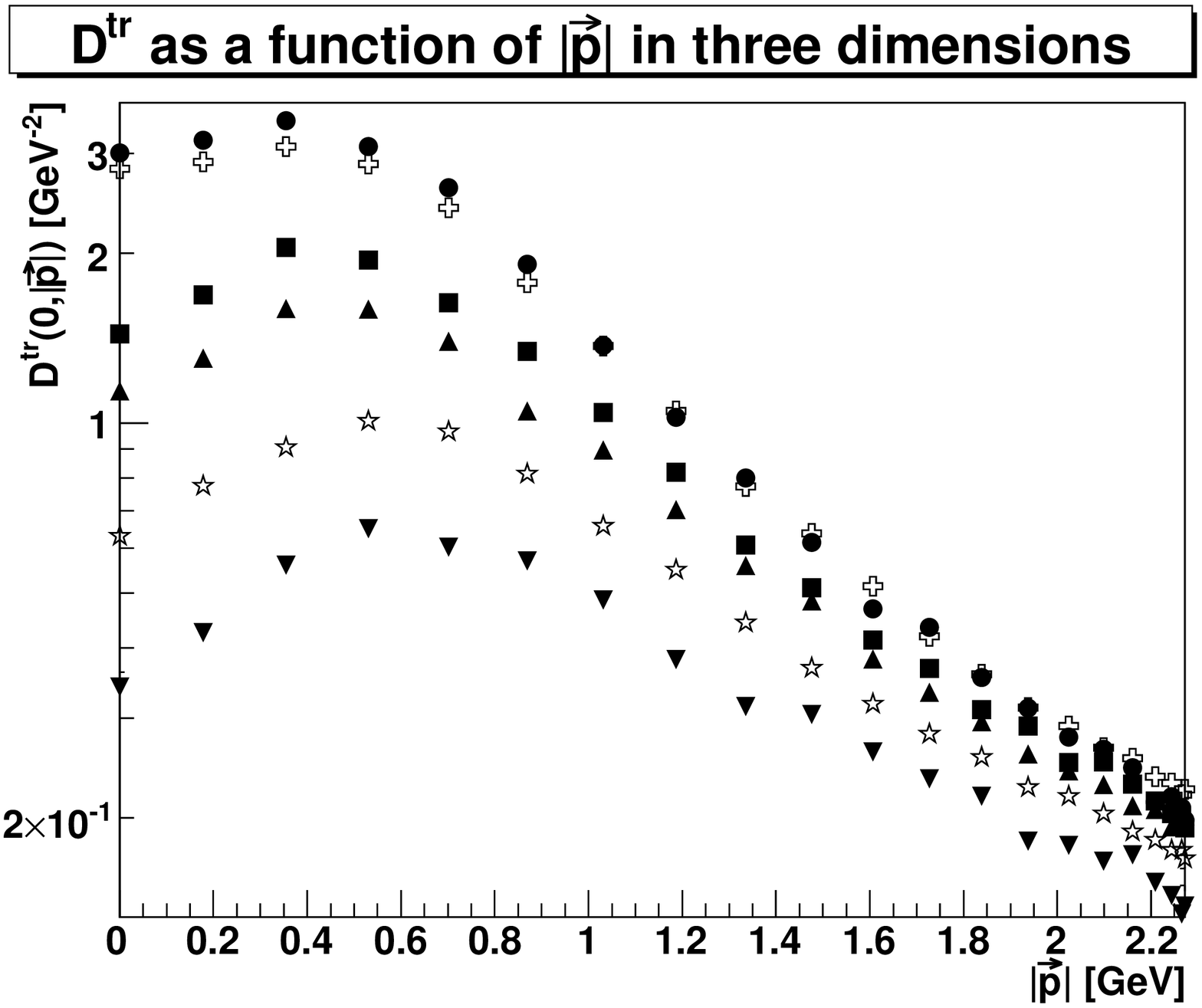}\\
\includegraphics[width=0.7\linewidth]{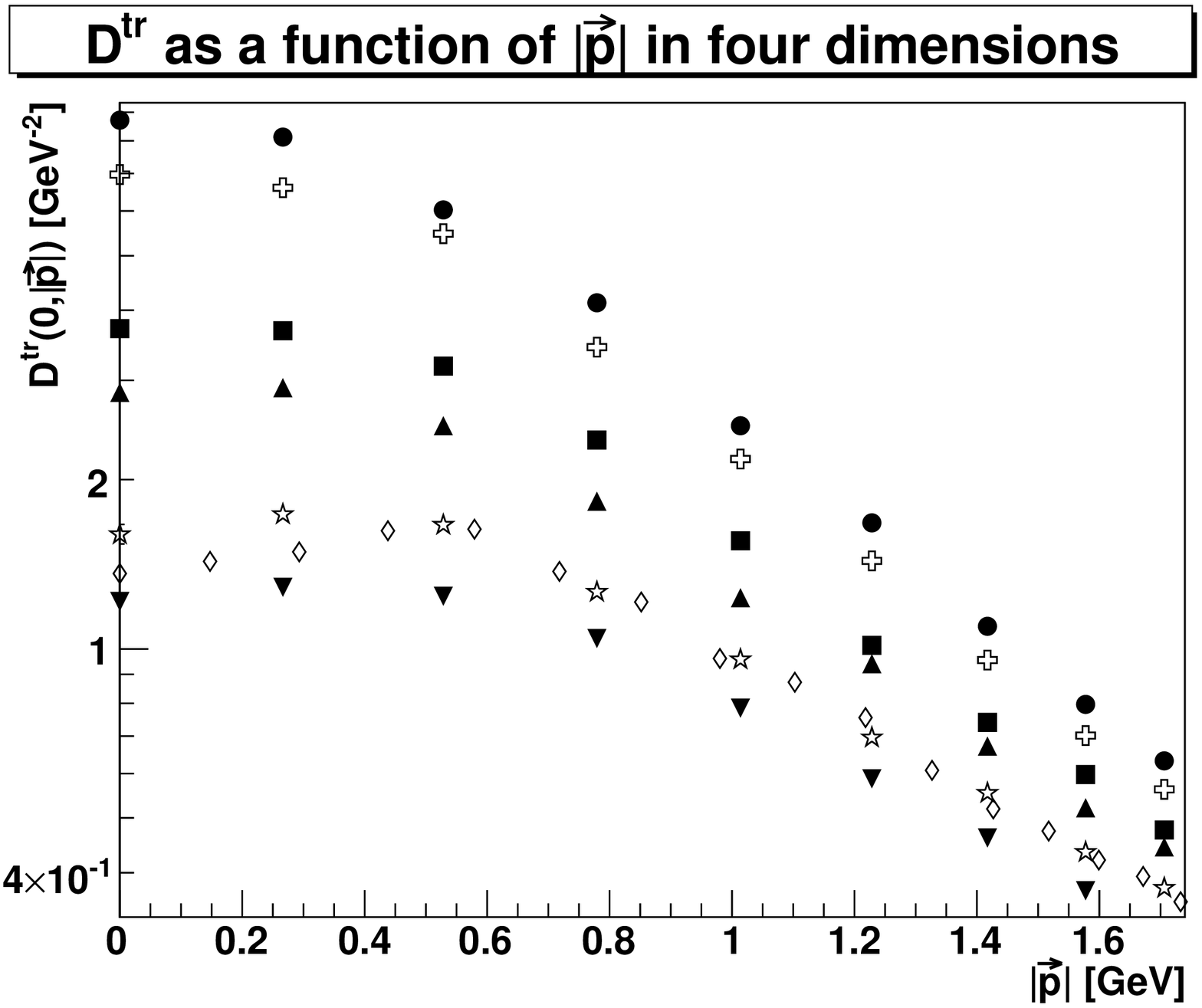}
\end{center}
\caption{\label{fig}
The gluonic correlation function $D^\tr(0,|\vec p|)$
in three (top figure) and four (bottom figure) dimensions.
The spatial momenta $|\vec p|$ are chosen along a single axis. Circles indicate data
for $\lambda=1$ (Landau gauge), crosses are used for $\lambda=1/2$, squares for
$\lambda=1/10$, triangles for $\lambda=1/20$, stars for $\lambda=1/100$ and
upside-down triangles represent results at $\lambda=0$ (Coulomb gauge). Data
have been obtained at $\beta=4.2$ in three dimensions and at $\beta=2.2$ in four dimensions.
The lattice size is $40^3 \approx $ (6.9 fm)$^3$ in three dimensions and $22^4 \approx$
(4.6 fm)$^4$ in four dimensions. In addition, in the bottom panel, diamonds correspond
to a $40^4 \approx $ (8.4 fm)$^4$ lattice at $\lambda=1/100$.
The physical scale has been set using Refs.~\protect\cite{Cucchieri:2003di,Fingberg:1992ju}.
Note that our error bars are smaller than the sizes of the symbols.
Also, for $\lambda=0$, we consider the standard (incomplete) Coulomb gauge.}
\end{figure}

\begin{figure}
\begin{center}
\includegraphics[width=0.7\linewidth]{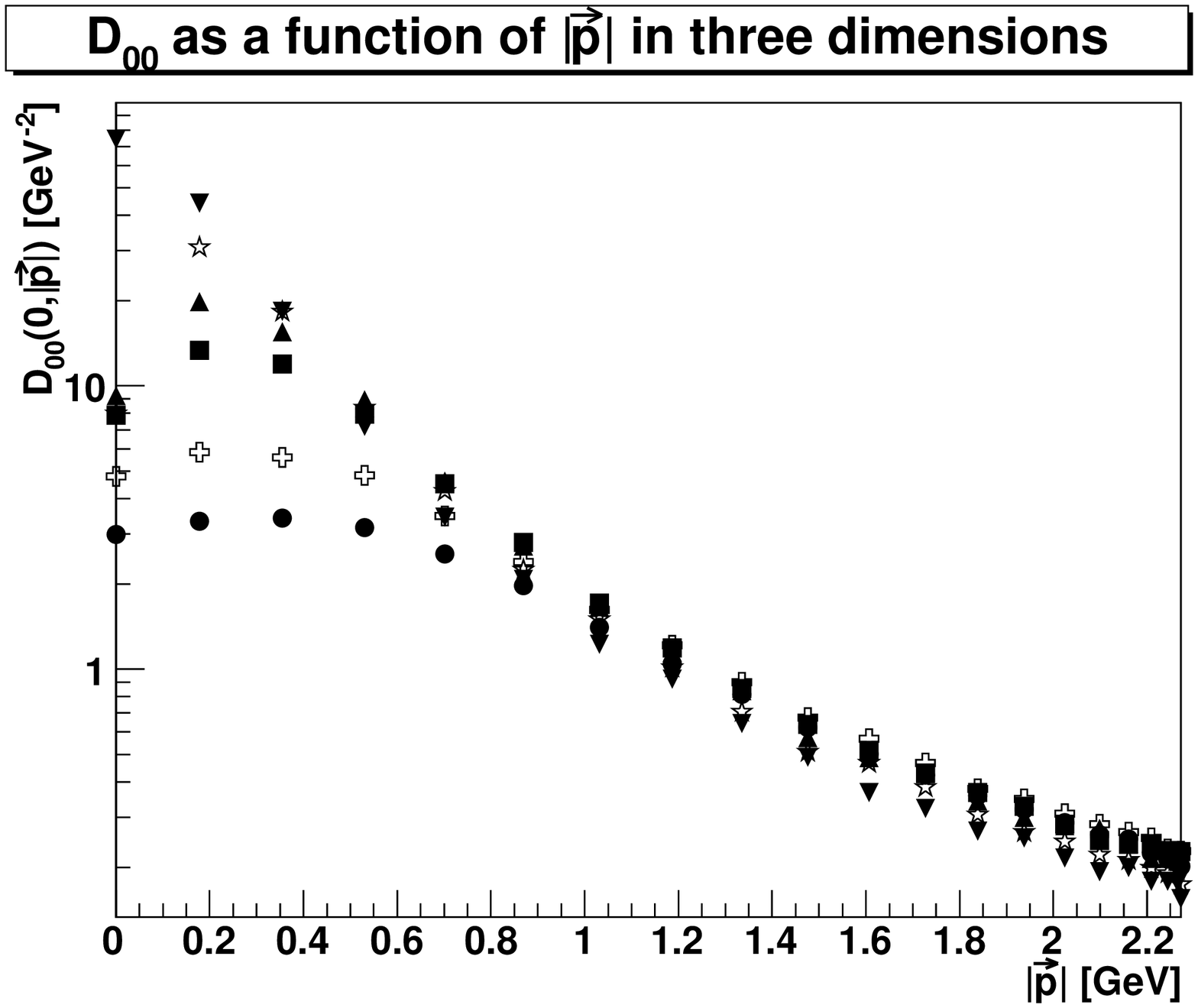}\\
\includegraphics[width=0.7\linewidth]{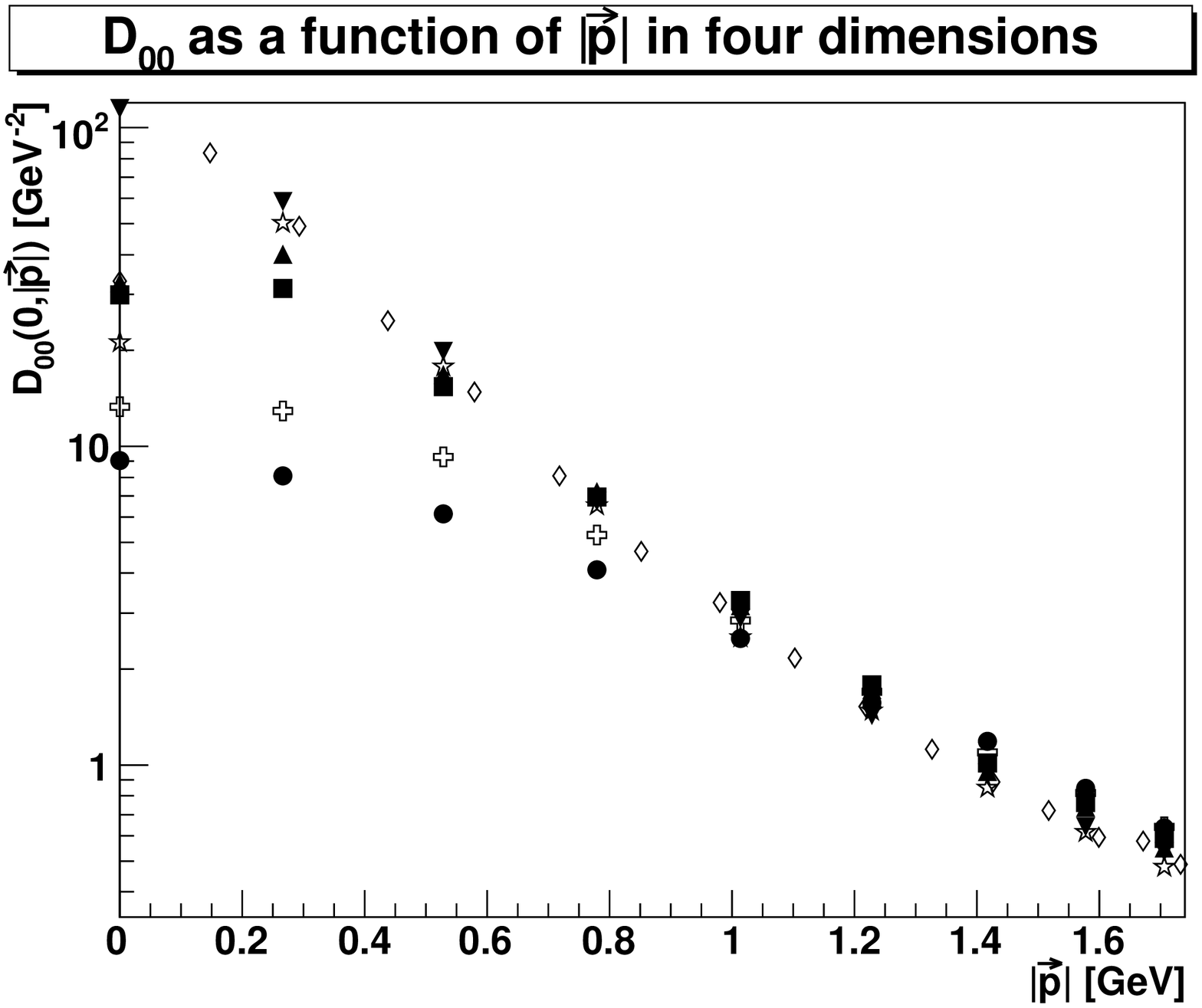}
\end{center}
\caption{\label{fig0}
Same as in Fig.\ \ref{fig}, except that here we show data for the $D_{00}(0,|\vec p|)$
tensor component.
}
\end{figure}

Our results\footnote{Preliminary results have been presented in Ref.~\cite{Maas,maas1}.}
are reported in Figure \ref{fig}. In three dimensions, 
a well-defined maximum (and thus an IR suppression) is visible
for all values of $\lambda$. This includes also Landau gauge ($\lambda = 1$),
confirming earlier results.\cite{Cucchieri:1999sz,Cucchieri:2001tw,Cucchieri:2003di}
As can be seen from the plot, the maximum value attained by the propagator seems
to move to larger momenta with decreasing $\lambda$, going from about 400 MeV in the 
Landau case\cite{Cucchieri:2003di} to about 600 MeV for the Coulomb case. 
At the same time, as $\lambda$ decreases, the maximum becomes more visible. 
In four dimensions, on the other hand, no discernible
peak is visible in Landau gauge when considering the volume $V = 22^4$.
Decreasing $\lambda$, however, leads to a suppression 
of the propagator in the IR region. In particular, at the smallest value of $\lambda$ 
considered, i.e.\ $\lambda=1/100$ and for a lattice volume $V = 40^4$,
a maximum is seen also in four dimensions.
This maximum is as visible as the one in three dimensions in Landau gauge. 
Thus for small $\lambda$, just as for Coulomb gauge,\cite{Cucchieri:2000gu}
one sees a maximum of the transverse gluon propagator already for relatively small 
volumes.

A similar $\lambda$-dependence is also seen in the tensor component $D_{00}(0,|\vec p|)$. In
particular, as $\lambda$ becomes small,
this tensor component of the gluon propagator becomes enhanced for small 
(nonzero) values of $|\vec p|$ (see Fig.\ \ref{fig0}). Let us recall that, in Coulomb gauge,
the $D_{00}$ component is IR divergent already at the perturbative level. Moreover,
the instantaneous part of $D_{00}$ is related to the color Coulomb potential
and should be diverging as $1/|\vec p|^4$ at small momenta.\cite{DanCoulomb,dc1,dc2,Cucchieri:2000gu,Cucchieri,cucchieri1,otherCoulomb,oc1}
Thus, from the discussion presented in the Introduction, one
could expect to observe a $D_{00}$ component enhanced at intermediate momenta $p > p_s$
but still suppressed at small momenta. This is indeed the case and from the plots
in Fig.\ \ref{fig0} one clearly sees how this enhancement is developing for $|\vec p| \neq 0$
as the value of $\lambda$ decreases. Let us stress that the discontinuity in the behavior
of the propagator at $|\vec p| \neq 0$ is observed since we are taking the limit
$\lambda \to 0$. A similar discontinuity is also obtained when implementing the complete
Coulomb gauge described in the Introduction.\footnote{This is a consequence
of the minimization of ${\mathcal E}_{\mbox{ver}}[g(t)]$, defined in Eq.\ \pref{Eresidual}.}
On the other hand, for the
incomplete Coulomb gauge, $D_{00}$ is clearly a continuous function of $|\vec p|$
(see Fig.\ \ref{fig0}).

As said in the Introduction, the IR exponent $\alpha_D$ for the transverse
correlator is predicted to show a discontinuity in the limit $\lambda\to 0$.
This is not seen from our data. However, a reliable check of this prediction
can only be obtained if one has control over the infinite-volume and the
continuum limits, which we have not yet achieved. On the other hand,
taking the limit $\lambda \to 0$ clearly induces a discontinuity in
the behavior of the temporal component $D_{00}$.\cite{Fischer:2005qe}


\section{Conclusions}

We have presented the first direct observation of an IR-suppressed gluonic correlation 
function on a symmetric 4d lattice in a Landau-like gauge, with gauge parameter
$\lambda$ as in Eq.\ \pref{gcinterpol}. (Landau gauge is obtained for $\lambda = 1$,
while Coulomb gauge corresponds to $\lambda = 0$.) 
The suppression is seen for sufficiently small $\lambda$ when
considering moderately small lattice volumes.
Judging from the results shown for the 3d case, it is conceivable that a similar suppression 
might be observed for any $\lambda$ if a large enough lattice side is considered.
Furthermore, since the limit $\lambda \to 1$ is smooth,\cite{Baulieu:1998kx,Fischer:2005qe}
we expect to see an IR suppression for sufficiently large lattices also in Landau gauge.
Moreover, we have obtained that the $D_{00}$ component of the gluon propagator gets enhanced
in the IR region (for $|\vec p| \neq 0$) at small values of the interpolating parameter
$\lambda$. These results clearly demonstrate the transformation of the correlation functions
from Landau gauge towards Coulomb gauge when considering the limit $\lambda \to 0$.
They also provide additional support to the Gribov-Zwanziger scenario of confinement,
establishing an IR suppression of gluonic correlation functions in four dimensions
beyond Coulomb gauge. The use of the interpolating gauge thus constitutes a
promising alternative to studies in Landau gauge, since finite-size effects
are significantly smaller, allowing a more efficient investigation of the gauge
dependence of correlation functions of confined objects.


\section*{Acknowledgments}

A.\ M.\ was supported in part by the DFG under grant number MA 3935/1-1.
A.\ C.\ and T.\ M.\ were supported by FAPESP (under grants \# 00/05047-5 and 05/59919-7)
and by CNPq. The ROOT framework\cite{Brun:1997pa} has been used in this project.
Part of our simulations has been done on the IBM supercomputer at S\~ao Paulo
University (FAPESP grant \# 04/08928-3).


\end{document}